# Matrix Inverse Free Method for Solving Quantum Electron Dynamics on Unstructured Grid

Katsuhiro Watanabe[1] and Akihito Kikuchi[2]

[1] Quantum programming institute, Cosmo-Stage-Wako 511, Chuo 2-4-3, Wako-shi, Saitama, 351-0113, Japan

[2] Advanced algorithm & systems, Dai-ni-Ito-Building 401, Shibuya-ku, Ebisu 1-13-6, Tokyo, 150-0013, Japan

Abstract

A matrix inverse free method to solve time-dependent Schrödinger equation is presented. The method is not subject to form of Hamiltonian and adopting real space grid system such as structured and unstructured grid, and achieves the order N algorithm even if we adopt unstructured grid systems to adjust complex regions. We have demonstrated some error evaluation problems that have exact solution and application in which it has non-Schrödinger type Hamiltonian, and we have successfully simulated the time-evolution of wave function for all of the case.

## I. Introduction

Understanding of electron characteristics as quantum dynamics is fundamental theme in all fields of physics and chemistry. In addition, it becomes also important in the engineering fields in relating to handle nano-technology recently. Especially ,when we consider (1)quantum dot and wire[1,2] ,(2)quantum chaos[3,4] ,(3)quantum tunneling[5] and (4)scattering[6], how to solve initial condition (IC) and boundary condition (BC) problem of time dependent Schrödinger equation (TDSE) is an important theme.

From the viewpoint of the numerical solution, this problem results in the proposition of the search of space discretization schemes and time integration algorithms. For space discretization, the methods are classified as real space methods such as finite deference method(FDM) ,finite element method(FEM) and solving the plane wave expansion with the aid of  fast Fourier transform(FFT).

Until now, it has been common way to make use of plane-wave expansion for space. It is still a mainstream in first-principle molecular dynamics and band calculation at present. It is true that the plane wave expansion method has the highest accuracy and resolution for identical orthogonal grid system. However, it can not be called the technique which is not always optimum concerning the calculation cost, when we treat the case in which wave function exists locally or there are coarse and dense distribution of potential field in computational region, and when we  mind parallel computing which is a trend in the age. There are several reports on real-space method, based upon such recognition. However, until now, real-space computations are executed only in orthogonal grid or body fitted coordinate(BFC), and the report which adopts the more flexible unstructured grid is not observed[7-13].

In the meantime, in the field of computational fluid dynamics (CFD), the problem of this kind is deeply tackled, and the rapid progress has already been accomplished. Proposal and selection of the scheme of the limitless number have been repeated. Following the recent studies of CFD, beyond the use of structured grid, such as (1)orthogonal or (2) BFC one, there is grid system optimization step in the unstructured grid system. In addition, CFD has the solution-adaptive-grid technology, which dynamically reconstructs the grids in the point where higher resolution is required [14]. There exist great numbers of idea which should be imported from CFD techniques, when electron wave-function must be traced in problems such as quantum chaos,  scattering and moving potential field situation in first principle molecular dynamics[14].

In CFD, spectral method, called as "plane wave expansion" in the field of computational physics, is no longer used except for the basic research of turbulent flow phenomena In addition, the recent research begins to have demonstrated that the performance of FDM, by adopting more than fourth order space accuracy, is equivalent to that of spectral method[15,16].  The search for real space methodology, if it is led by the knowledge of CFD, can be called as a trend in next generation in the computational physics.

It seems that the real space methods, proposed until now, have not been completed yet.  For example, the differential operator splitting approach cannot be applied, when one aims at the unstructured grid. Meanwhile, the multi-step method is explicit method, so it performs on unstructured grid. But, explicit time integration method does not conserve the norm of wave function generally.  For this problem, Iitka[7] proposed a solution by adopting higher order time integration approach. However, it is desirable that the time integration is essentially unitary, when practical and complicated problems are handled and considered. In short, the symmetrical treatment in the time-direction, as is represented by Crank-Nicolson method, becomes indispensable in solving TDSE. It means that the inverse problem of sparse matrix should be solved in the unstructured grid system. In the solution method of the sparse matrix, the problem of fill-in arises, when direct method was adopted, and it is not desirable because many machine memories are required. In the iteration methods such as conjugate gradient method, extra work area is required and in addition, the total amount of the numerical operation increases because of the requirement for the repeated calculation.

Concerning with these problems, we propose a scheme that makes Hamiltonian into discrete one at first and next decomposes the exponential operator into LDU form. By this scheme, we have succeeded in the development of high-speed, grid-system-adoptive and memory-saving computational method, which is free from matrix calculation.
This method is the order N algorithm which is perfectly proportional to number of non-zero element of discretized Hamiltonian matrix.

In section 2, universality and superiority of present approach namely "Matrix Inverse Free Method" is discussed. In section 3, several problems in which the exact solution exists are taken up, and the error evaluation is carried out. In section 4, the simple application to non-Schrödinger type differential operator is demonstrated. In last section, some problems of present method are arranged as concluding with future prospect.

## II. Formulation

### A. Concept : Path of Solving Process

When there is no time-dependence term in Hamiltonian, the formal solution of wave function can be written as follows.

$$\psi(\mathbf{x},t) = \exp(-itH)\psi(\mathbf{x},0) \qquad (1)$$

Two paths will be available as the procedure for obtaining numerical solution based on this formal solution.

(1) Path Type I
Step1) The solution is formally obtained, using the differential operator expression of Hamiltonian.
Step2) The exponential operator is decomposed into simple operator product.
Step3) Decomposed exponential operator is expanded as Tayler series.

Step4) The expanded differential operator is finally discretized by a certain way, such as difference calculus.

This approach is the method proposed by Watanabe *et al*. [10,11].

(2) Path Type II

Step1) First, Hamiltonian is discretized, and it is converted into simultaneous ordinary differential equations on the finite dimensional wave function vector,

$$\frac{d\mathbf{\psi}}{dt} = i\mathbf{H}\mathbf{\psi}. \tag{2}$$

In the above, $\mathbf{\psi} = (\psi_1, \psi_2, \cdots, \psi_N)^T$ is approximation by N-dimensional vector for wave-function and $\mathbf{H}$ is discrete representation of Hamiltonian operator respectively.

Step2) The formal solution of ordinary differential equation on vector form is obtained as,

$$\mathbf{\psi}(t + \Delta t) = \exp(-i\Delta t \mathbf{H})\mathbf{\psi}(t). \tag{3}$$

Here, "Δt" means time step interval.

Step3) Superscript matrix of the exponential function is decomposed.

Step4) Finally, each of decomposed exponential matrix operators is expanded with respect to Δt.

The latter path is the approach which De Raedt[13] and our present work have adopted. The difference of these two approaches is whether or not the discretization is executed first. However, it leads two approaches to decisively different style.

B．Algorithm : Matrix Inverse Free Method

Generally, it is possible to decompose matrix $\mathbf{H}$ in lower, diagonal and upper form as a triangular matrix,

$$\mathbf{H} = \mathbf{L} + \mathbf{D} + \mathbf{U}. \tag{4}$$

Therefore, formal solution after the space discretization i.e., $\mathbf{\psi}(t + \Delta t) = \exp(-i\Delta t \mathbf{H})\mathbf{\psi}(t)$ is approximately decomposed with as follows.

$$\begin{aligned}\mathbf{\psi}(t + \Delta t) &= \exp(-i\Delta t \mathbf{H})\mathbf{\psi}(t) \\ &= \exp\{-i\Delta t(\mathbf{L} + \mathbf{D} + \mathbf{U})\}\mathbf{\psi}(t) \\ &\cong \exp(-i\Delta t \mathbf{L})\exp(-i\Delta t \mathbf{D})\exp(-i\Delta t \mathbf{U})\mathbf{\psi}(t)\end{aligned} \tag{5}$$

The lower and upper triangular matrix will be redefined by the consideration of the simplicity of the treatment in actual calculation,

$$\widetilde{\mathbf{L}} = \mathbf{L} + \frac{1}{2}\mathbf{D}, \quad \widetilde{\mathbf{U}} = \mathbf{U} + \frac{1}{2}\mathbf{D}, \tag{6}$$

and, when we consider unitality and second degree time accuracy, each of exponential matrix operators must be ordered as follows,

$$\psi(t+\Delta t) = \exp\left(-\frac{1}{2}i\Delta t\widetilde{\mathbf{L}}\right)\exp\left(-\frac{1}{2}i\Delta t\widetilde{\mathbf{U}}\right)\exp\left(-\frac{1}{2}i\Delta t\widetilde{\mathbf{U}}\right)\exp\left(-\frac{1}{2}i\Delta t\widetilde{\mathbf{L}}\right)\psi(t). \tag{7}$$

In addition, it is calculated in the following way for the sake of keeping the symmetry of the treatment.

$$\psi(t+\Delta t) = \frac{1}{2}\left\{\begin{array}{l} \exp\left(-\frac{1}{2}i\Delta t\widetilde{\mathbf{L}}\right)\exp\left(-\frac{1}{2}i\Delta t\widetilde{\mathbf{U}}\right)\exp\left(-\frac{1}{2}i\Delta t\widetilde{\mathbf{U}}\right)\exp\left(-\frac{1}{2}i\Delta t\widetilde{\mathbf{L}}\right) \\ +\exp\left(-\frac{1}{2}i\Delta t\widetilde{\mathbf{U}}\right)\exp\left(-\frac{1}{2}i\Delta t\widetilde{\mathbf{L}}\right)\exp\left(-\frac{1}{2}i\Delta t\widetilde{\mathbf{L}}\right)\exp\left(-\frac{1}{2}i\Delta t\widetilde{\mathbf{U}}\right) \end{array}\right\}\psi(t)$$

(8)

This computation is reduced to two types.

$$\psi_L = \exp\left(-\frac{1}{2}i\Delta t\widetilde{\mathbf{L}}\right)\psi_{LorU} \tag{9}$$

$$\psi_U = \exp\left(-\frac{1}{2}i\Delta t\widetilde{\mathbf{U}}\right)\psi_{LorU} \tag{10}$$

These are approximated by the first order Tayler expansion of which the conjugation is symmetrical on the time direction so that the norm of the wave function should be preserved.

$$\left(\mathbf{I}+\frac{1}{4}i\Delta t\widetilde{\mathbf{L}}\right)\psi_L = \left(\mathbf{I}-\frac{1}{4}i\Delta t\widetilde{\mathbf{L}}\right)\psi_{LorU} \tag{11}$$

$$\left(\mathbf{I}+\frac{1}{4}i\Delta t\widetilde{\mathbf{U}}\right)\psi_U = \left(\mathbf{I}-\frac{1}{4}i\Delta t\widetilde{\mathbf{U}}\right)\psi_{LorU} \tag{12}$$

The calculation for the unknown in the left hand side is explicitly obtained by sweeping forward from the first index for $\widetilde{\mathbf{L}}$ matrix and sweeping backward from the maximum index for $\widetilde{\mathbf{U}}$.

That is to say, all calculation is completed only in the explicit treatment of forward and backward sweeping of matrix $\widetilde{\mathbf{L}}$ and $\widetilde{\mathbf{U}}$, and the scheme, which completely does not require the inverse matrix calculation, can be constituted.

### C. Merits of Present Scheme

For Universality :
1. The scheme is not subject to the change on the space dimension.

2. The scheme is not subject to the change on the lattice system such as structured or unstructured grid.
3. The procedure is not dependent on the concrete form of kinetic and potential term in Hamiltonian representations.
4. Present scheme is able to treat arbitrary periodic problem by ghost cell techniques.

For Computational Cost :

In case of the calculation with second order special accuracy in structured grid, since only tri-diagonal matrices are generated by the expansion in xyz axes, the direct calculation is possible by the disorption method[10,11]. Therefore, the inverse matrix calculation is substantially unnecessary. However, such an advantage in this approach is lost in the unstructured grid systems. In this case, the use of the linear solver for the sparse matrix is required in the simplicity. For sparse matrix, direct methods are not desirable because the "fill-in" is occurred. This leads machine memory increasing. In the meantime, the total amount of the numerical operation.

As for the present algorithm, even in such cases, the inverse matrix calculation is not required and the calculation is possible with the computational cost in proportion to the number of the off-diagonal non-zero elements.

## III. Error Evaluation

In this section, (A)the problem of diffusion process and (B)the motion problem of wave packet is taken up as error evaluation of this technique, and it is compared with each exact solution, and it is shown to be the equivalent performance with past methods in case of the simple structured grid systems.

### A. Diffusion Problems

In imaginary time direction, time-dependent Schrödinger equation's problems are analogues to initial and boundary condition problems in classical diffusion process. This is as follows,

$$\frac{\partial \psi}{\partial \tau} = H\psi \quad . \tag{13}$$
$$\tau = -it$$

For simple discussion , we consider free particle problem. In this case, Hamiltonian is equal to Laplacian $\nabla^2$, and represented as,

$$\nabla^2 = \frac{\partial^2}{\partial x^2} + \frac{\partial^2}{\partial y^2} + \frac{\partial^2}{\partial z^2}, \tag{14}$$

in Cartesian coordinates. It is compared with the exact solution in order to evaluate the error of the present technique. On the evaluation of the problem, see Abuduwali *et al.* of the reference[17].

### Case1   One-Dimensional Exact Solution

Initial Condition is set as follows,

$$\psi(x,0) = 100\sin(\pi x) \quad for \quad x \in [0,1] \tag{15}$$

and boundary value is set to zero at x=0 and x=1. Exact solution of this condition is given as following form.

$$\psi(x,\tau) = 100\exp(-\pi^2 \tau)\sin(\pi x) \tag{16}$$

The calculated result for spatial distribution is shown in figure1(a), relative error distribution is shown in figure1(b). In figure1, a time step parameter "$\alpha$" is defined as $\alpha = \Delta t \times \max(abs(D_i)), i = 1,\ldots N$, and, $D_i$ is diagonal element of I-th row.

In figure1(b), it is observed that maximum error exists at neighbor of fixed boundary and minimum error exists at middle point. One of the reasons will be as follows. Non-zero values of commutation relation between $\tilde{\mathbf{L}}$ and $\tilde{\mathbf{U}}$ i.e., $[\tilde{\mathbf{L}}, \tilde{\mathbf{U}}] \equiv \tilde{\mathbf{L}}\tilde{\mathbf{U}} - \tilde{\mathbf{U}}\tilde{\mathbf{L}}$ are the origin of the numerical errors in the splitting of operators. These non-zero elements are located only on the off-diagonal of the row that corresponds to the grid points in neighbor of fixed boundary points, and the numerical errors concentrate there. But, such situation is not critical problem for electron dynamics computation because we are almost interest in domain wave function distribution area   From these, it is verified that good approximate solution can be obtained by present algorithm.

### Case2   Two-Dimensional Exact Solution

Initial and boundary conditions are similar to one-dimensional problem.

$$\psi(x,y,0) = 100\sin(\pi x)\sin(\pi y) \quad on \quad \Omega(x,y)$$
$$\Omega(x,y) = [0,1] \times [0,1] \tag{17}$$

$$\psi(x,y,\tau) = 0 \quad on \quad \partial\Omega \tag{18}$$

In this case, exact solution will be given as following.

$$\psi(x,\tau) = 100\exp(-2\pi^2 \tau)\sin(\pi x)\sin(\pi y) \tag{19}$$

The unstructured grid system used for the calculation is the array of square cell, each of which is furthermore split into two right-angled triangles as the base element( see figure2).

The finite element method was adopted for the discretization of Hamiltonian, and the linear element is used as the basis function. In present work, the mass matrix, which arises with respect to the time term, was diagonalized by lumped mass approximation. After the scaling of both sides in the equations by the diagonalized mass matrix, the exponential operator expression is constructed and the LU decomposition method was applied. Result of the calculation are shown at table I. Details of this computation condition, for example grid size, time interval, time integration counts, and so on, are same as the reference[17].

In table I, the definition of "$\alpha$" had been already mentioned in the previous, the "Error" is estimated by $Error = (f_{computaion} - f_{exact})/f_{exact}$ , and "Local Cranl-Nicolson method" that has been presented by Abuduwali *et al* [17] is almost equivalent to the method that Watanabe *et al.* have proposed for 2-dimensional case[10,11]. It is verified that, even in this case, the present algorithm maintains the accuracy of the same grade as Crank-Nicolson method, which is the standard solution method of problem of this kind.

B．Wave Packet Problem

The problem by Watanabe *et al.* is taken up[10]. In this problem, all of computational conditions are exactly the same that the other worker has adopted. The details are given in the literature of reference[10]. It is obvious that the present work gives the result that is similar to the other works, as is shown in table II. Also present work gives well approximated values in comparison with exact values.

## IV. Application

The LDU decomposition scheme for quantum electron dynamics proposed in the present work can be applied to various systems, in which the Hamiltonian is not necessarily expressed by the differential operator of Schrödinger type. For example, it will be applicable to the case where the equation for the "effective" Hamiltonian is given as the following form,

$$i\frac{\partial \psi}{\partial t} = \varepsilon(\vec{p})\psi = \varepsilon(-i\vec{\nabla})\psi \quad , \tag{20}$$

and the suitable finite-difference representation is given. In addition, the LDU decomposition scheme can be applicable to the case in which the Hamiltonian is expressed as "tight-binding" matrix form. In this section, an example of this is given. We consider a model described by a bracket Hamiltonian including multi-channel scattering between the two states "e" and "h" as follows.

$$H = \sum_l \left( T|l;e\rangle\langle l+1;e| + T|l+1;e\rangle\langle l;e| \right)$$
$$+ \sum_l \left( -T|l;h\rangle\langle l+1;h| - T|l+1;h\rangle\langle l;h| \right). \tag{21}$$
$$+ \sum_i \Delta(l)\cdot\left( |l;e\rangle\langle l;h| + |l;h\rangle\langle l;e| \right)$$

In the above, the states "e" ("h") defined on $l$-th atomic site is expressed as $|l;e(h)\rangle$. The parameters T and $\Delta(l)$ are the hopping integral and the pair potential between states "e" and "h" on l-th site. We can trace the time-evolution of the wave-function expressed as vector form, spanned by $\langle\phi|l;e\rangle$ and $\langle\phi|l;h\rangle$, making use of matrix representation of the Hamiltonian $\langle l|H|l'\rangle$

In the uniform system, i.e., $\Delta(l)=\Delta$ on all sites, the energy band is given as

$$E = \pm\sqrt{4T^2\cos^2(ka) + \Delta^2} \quad , \tag{22}$$

by means of Bloch-function of following form,

$$\langle k^{e(h)}|l;e(h)\rangle = \frac{1}{\sqrt{N}}\exp(i\cdot k^{e(h)}\cdot l\cdot a). \tag{23}$$

In the above, parameter "a" means the distance of the nearest atoms. In the crystal with the band structure as above, there is no state with energy E, such as $|E|<|\Delta|$ or $|E|>|2T|$. Hence, only the electron waves of $|\Delta| \leq |E| \leq |2T|$ can propagate there. Even if electron wave with energy E, such as $|E|<|\Delta|$, is injected from the outside of the crystal, it will be reflected backwardly.

We have successfully simulated the time-evolution of the wave packet in a junction system, where the pair potential $\Delta(l) \neq 0$ acts in a "scattering" zone of finite length. The detailed condition for the calculation is shown in the table III. In this case, due to the presence of pair potential, the energy gap at E～0 exists locally in the scattering zone. The initial wave packet, lying channel "e" only, is taken to be the following form,

$$\psi_{\_e}(n) = \frac{1}{(2\pi W^2)^{\frac{1}{4}}} \exp\left(\frac{-(n \cdot a - n_0 \cdot a)^2}{4W^2} + i \cdot p_0 \cdot n \cdot a\right), \quad (24)$$

at $n$-th atomic site, so that the main part of the initial packet is located outside of the scattering zone. Hence the initial wave-function is assumed to lie in the part of the crystal where the band structure is given as $E = \pm 2T\cos(ka)$, since $\Delta(l)$ is zero there. From the viewpoint of the energy band, the wave-packet is emitted toward the "energy gap" of the scattering zone, since the packet momentum $p_0$ is set to be $1/2(\pi/a)$. Figures show the calculated result of the time-evolution of the wave packet. After 2000 time steps, the wave packet is split into two parts, which separately go in the opposite directions(Figure 3). The behavior of the packet can be explained by the discussion of "band gaps" as above. The left-going packet mainly consists of the partial wave, which justly injects into the energy gap of the scattering zone, being reflected backward by the energetic reason. On the other hand, the right-going packet is the set of waves, which tunnel through, or, go outside of, the "energy gap" in the scattering zone. In the figure 4, which shows the k-space FFT spectrum of the wave-function, such a situation can be seen as the depression at k～$1/4(\pi/a)$ of the plotted spectrum for wave-function of state "h" and also as the peak of that for state "e" at k～$3/4(\pi/a)$. Throughout the simulation, the norm and the energy of the wave-function is well conserved also in this example.

## V. Conclusion

In conclusion, we have developed a new order N algorithm for the time-dependent Shrödinger equation that is the *Matrix Inverse Free Method*. By this algorithm, explicit time integration procedure has become possible in spite of the fact that implicit time-discretization formulation must be adopted to conserve the norm of wave function. Also, this algorithm has invariance for utilizing spatial grid system and Hamiltonian type.

In this article, we have briefly described the basic idea of the method by illustrating performance with evaluating problem in which it has exact solution and numerical application to multi-channel scattering problem. From these demonstrations, we have found that the present algorithm is quite efficient for simulating time-evolution of wave function. But also we have found that approximation error increases nearby the boundary on which Diriclet condition is cast. This error may be critical for the case of quantum chaos simulation because the reflection of wave function at infinite potential wall is key phenomenon. For this problem, reducing of the error will become possible by adopting higher accurate discretization scheme for time and spatial directions. We will improve and resolve in future work.

Table I.

The comparison of the exact solution and numerical solutions at middle point value with $\alpha=4$, $\tau=0.1$.

|  | Value at x=y=0.5 | Error |
|---|---|---|
| Exact solution | 13.891119 |  |
| Crank-Nicolson method | 14.095628 | $1.472\times10^{-2}$ |
| Local Crank-Nicolson Method | 14.028540 | $9.827\times10^{-3}$ |
| Present method | 13.959336 | $4.887\times10^{-3}$ |

Table II.

The comparison of the exact solution and numerical solutions for one-dimensional wave packet motion problem.  In second column, error of wave function's norm are shown, $\langle p(t) \rangle, \langle E(t) \rangle$ and $\langle x(t) \rangle$ mean expectation value of momentum, energy and center point of wave packet respectively.

|  | $abs\left(\int_{\Omega} |\Psi| dx - 1.0\right)$ | $\langle p(t) \rangle$ | $\langle E(t) \rangle$ | $\langle x(t) \rangle$ |
|---|---|---|---|---|
| Exact | 0 | 12.0 | 72.0 | 3.19532 |
| Watanabe *et.al.* | $< 1.0^{-7}$ | 11.69785 | 73.01940 | 3.17048 |
| Present work | $< 1.0^{-7}$ | 11.69785 | 73.01940 | 3.16647 |

Table III.

Parameters for the calculation is given here. Units are taken to be dimensionless and scaled by hopping integral $T$ and the distance of nearest atoms $a$.

|  | Hopping integral | Pair potential $\Delta(l)$ | Width |
|---|---|---|---|
| Left asymptotic zone | $T\ (<0)$ | 0 | $\approx 500a$ |
| Middle scattering zone | $T$ | $\frac{1}{10}T$ | $32a$ |
| Right asymptotic zone | $T$ | 0 | $\approx 500a$ |

| Time step $\Delta t$ | $\Delta t \cdot |T| = \frac{1}{8}$ |
|---|---|
| Packet momentum $p_0$ | $\frac{1}{2}\frac{\pi}{a}$ |
| Packet width $W$ | $8a$ |
| Width of calculation space | $1024a$ |

# Figure Captions

Figure 1. Comparison of exact solution and the present numerical solutions for several $\alpha$ values with $\tau$ =0.5. (a) Spatial distribution of wave function (b) Spatial distribution of normalized error that is defined as in the figure.

Figure 2. Illustration of mesh system used in 2-dimensional evaluation problem. Distance is dimensionless.

Figure 3. Time evolution of the wave-packet is shown. The wave-functions for the states "e" and "h" are denoted as $\phi\_e$ and $\phi\_h$, respectively. The amplitudes of the initial and final wave-function (after 2000 time steps) are plotted here. In the figure, the scattering region, where the pair potential is acting, is shown as the shaded zone.

Figure 4. Fourier spectrum of the wave-function, which is defined as $W(k) = \frac{1}{N}\sum_{n}\exp(i \cdot k \cdot na) \cdot \psi(n)$, is shown. ( The amplitude of $W$(k) is plotted here.) The depression of the plotted spectrum |W(k)| for wave-function $\phi\_h$ (final) at k~1/4($\pi$/a) and the peak of that for $\phi\_e$(final) at k~3/4($\pi$/a) result from the scattering mechanism as discussed in the text. Since partial waves of k~1/4($\pi$/a) in the initial packet enter into the local "energy gap", they are strongly scattered. Some part of them are reflected backwardly and induce the peak of plotted spectrum at k~3/4($\pi$/a).

Figure 1.

(a) Wave function

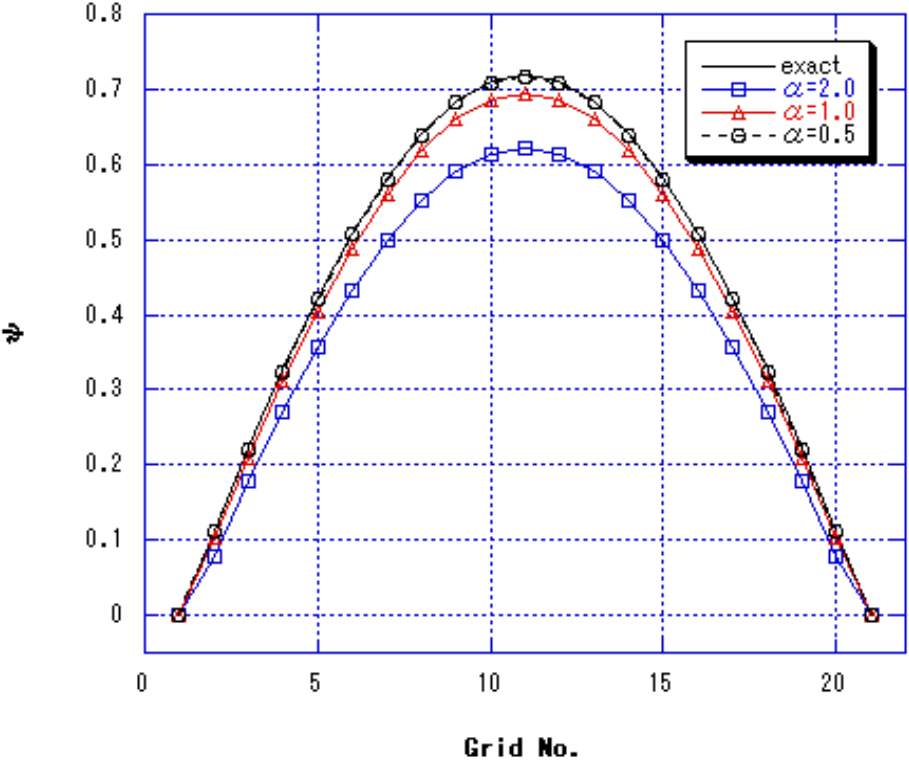

(b) Normalized error

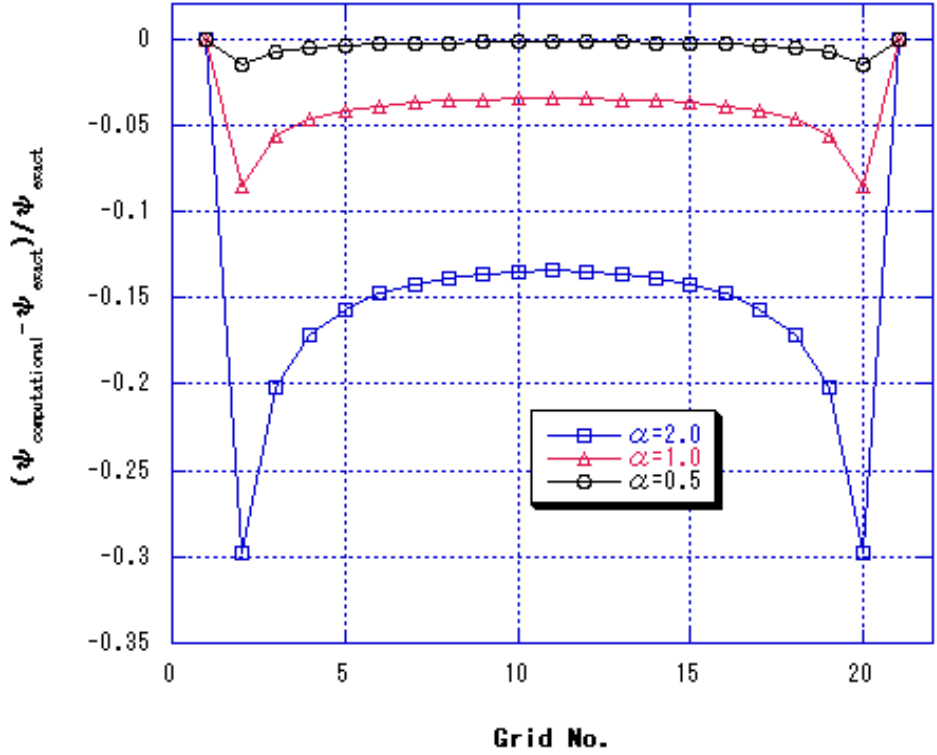

Figure 1. Comparison of exact solution and the present numerical solutions for several $\alpha$ values with $\tau=0.5$. (a) Spatial distribution of wave function (b) Spatial distribution of normalized error that is defined as in the figure.

Figure 2.

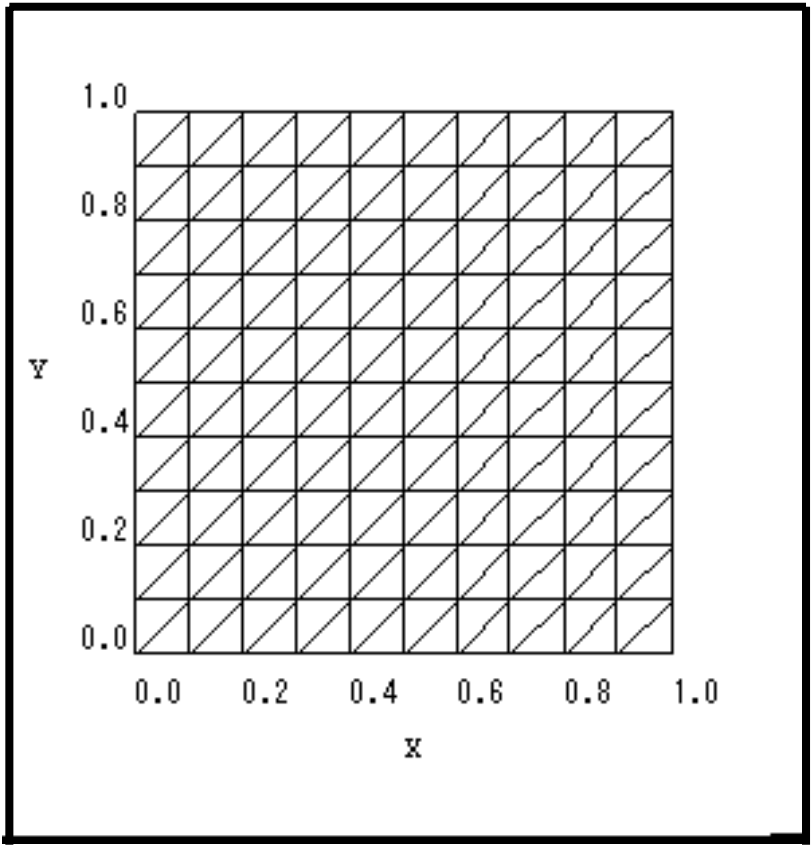

Figure 2.　Illustration of mesh system used in 2-dimensional evaluation problem. Distance is dimensionless.

Figure 3

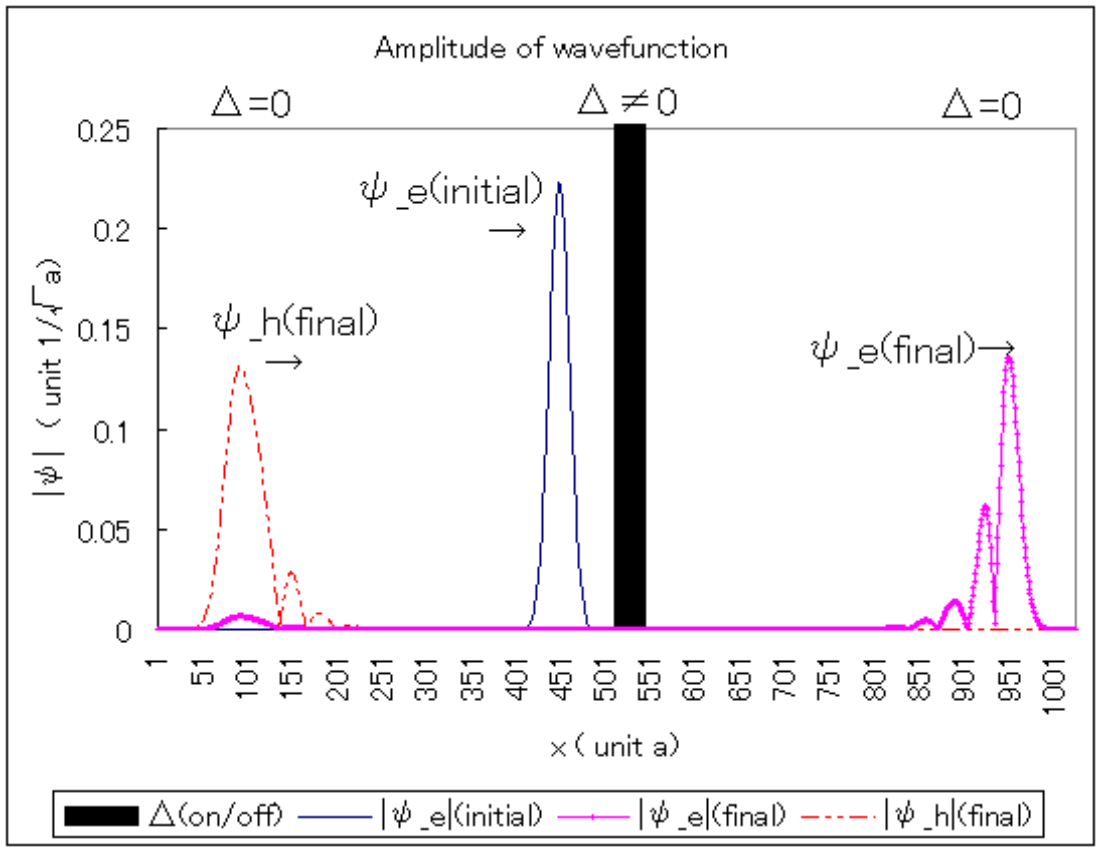

Figure 3. Time evolution of the wave-packet is shown. The wave-functions for the states "e" and "h" are denoted as $\phi\_e$ and $\phi\_h$, respectively. The amplitudes of the initial and final wave-function (after 2000 time steps) are plotted here. In the figure, the scattering region, where the pair potential is acting, is shown as the shaded zone.

Figure 4.

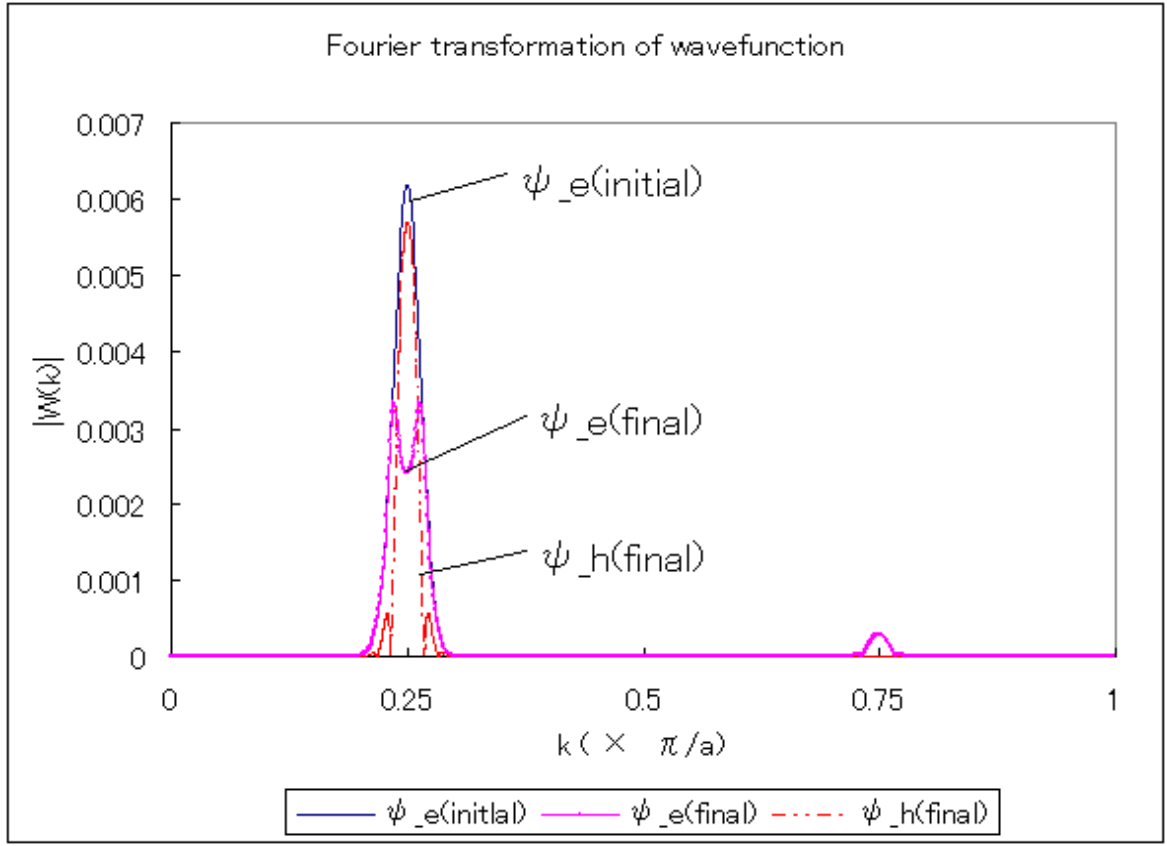

Figure 4. Fourier spectrum of the wave-function, which is defined as $W(k) = \frac{1}{N} \sum_n \exp(i \cdot k \cdot na) \cdot \psi(n)$, is shown. ( The amplitude of W(k) is plotted here.) The depression of the plotted spectrum |W(k)| for wave-function ψ_h (final) at k〜1/4(π/a) and the peak of that for ψ_e(final) at k〜3/4(π/a) result from the scattering mechanism as discussed in the text. Since partial waves of k〜1/4(π/a) in the initial packet enter into the local "energy gap", they are strongly scattered. Some part of them are reflected backwardly and induce the peak of plotted spectrum at k〜3/4(π/a).